# Quantitative Survey on Extreme Programming Projects

**Bernhard Rumpe**
Software & Systems Engineering
Munich University of Technology
Arcisstr. 21
D-80333 Munich, Germany
+49 89 289 28129
Bernhard.Rumpe@in.tum.de

**Astrid Schröder**
Software & Systems Engineering
Munich University of Technology
Arcisstr. 21
D-80333 Munich, Germany
+49 89 289 28129
Astrid.Schroeder@in.tum.de

**ABSTRACT**
In recent years the Extreme Programming (XP) community has grown substantially. Many XP projects have started and a substantial amount are already finished. As the interest in the XP approach is constantly increasing worldwide throughout all software intensive application domains, it was time to start a first survey on XP. This paper presents the results of 45 evaluated questionnaires that have been received during the Summer 2001 survey.

## 1 INTRODUCTION

Based on a series of books and a large body of conference, magazine, and journal papers, the Extreme Programming approach to software development is widely known and has become rather prominent. A number of pilot projects using the XP approach has been started. However, many companies are still facing the question, whether, in which projects and in which form they should move from their traditional or object-oriented approaches to software development to Extreme Programming. Supporters of XP claim a larger number of benefits, but today statistical quantitative support for these claims has not been given. As XP exists for a number of years, it is time to start gathering data.

This article describes a survey based on 45 questionnaires, which was conducted during Summer 2001. In Section 2, we describe the content of the survey, and how people have been addressed. In Section 3, we present a condensed version of the survey results and give a final outlook in Section 4.

For those interested in an introduction to or further reading on XP, we recommend [1,2,3,4] or more scientific articles in the proceedings [5] that contain e.g. [6]. The full study is available as [7].

## 2 STRUCTURE OF THE SURVEY
**The questionnaire**
The purpose of this survey was to get a general understanding of the current situation in XP projects, the problems, the kind of projects using the XP approach, the results, the background of the team members etc. As XP people are typically busy, we decided not to ask all interesting questions, but to concentrate on three blocks of total 33 questions. The questions are:

**Block 1. On the Company**
1,2: Name of project, person, company are not disclosed, but were collected for possible additional questions and to prevent several questionnaires on the same project.
3. Role of person who filled in this questionnaire
4. City and country where company is located
5. Some information about the company (how big, founded when, what line of business is it in, how many other XP projects were carried out before?, ...)

**Block 2. On the XP-Project**
6. Duration of project (from when till when)
7. Team size
8. Total manpower e.g. in person-months
9. How good was the general software engineering training/knowledge of the team members initially?
10. How many team members had made experiences in XP previously?
11. How many development companies/independent consultants were involved?
12. Why did you decide to develop this project with XP?
13. Programming languages used
14. Technologies used
15. What kind of software was developed?
16. Has it been a development from scratch (new system), legacy maintenance, or adding new functionality on an existing system?
17. What was the project structure (how many people were there for each role)?
   - Programmers (writing production code and code for component tests)
   - Customers
   - Testers (helps customer developing functional tests)
   - Coach
   - Further roles (consultant, big boss, tracker...)

18. How many customers with different stakes (requirements, forms of usage for the system) were involved?
19. Did the project terminate successfully? (9=very successful, 0=not at all successful)
20. What were the major reasons for its success / failure? Can you priorize them
21. If it was a success what were the main obstacles? How dangerous have they been?
22. XP was invented to make software development more successful. Some of its main goals are listed below. In your XP-project, could these goals be reached? If not, explain the obstacles? (5=fully achieved, 0=as always, -5=much worse)
    - Deliver software in time
    - Let developers have fun with their work
    - Develop software with a high quality (less bugs)
    - Late changes don't cause high costs, because one can react fast to changes
23. Which XP-Elements did you use in the Project? (9=fully used, 0=not at all) Please say for every element how strong you used it (9-0) and if you consider it helpful(h), improvable(i), or even making-difficult(m) for success of development.

    | Planning Game | Pair Programming |
    | Short Release Cycles | Common Code Ownership |
    | Metaphor | Continuous Integration |
    | Simple Design | 40-Hour-Week |
    | Testing | On-Site Customer |
    | Refactoring | Coding Standards |

24. Please give reasons for the three least used concepts, why you didn't use them? Did you explicitly decide not to, or had there been other obstacles?
25. Do you have improvement suggestions for any of the XP elements (perhaps in your project you already used this elements in the way you improved it for yourself)?
26. Have you used additional concepts, tools or modeling languages that go beyond the pure XP approach? How did they integrate to XP?
27. Some comments about the project and the project progress
28. Further comments

**Block 3. Future plans and personal background**
29. Will you use XP again?
30. Are you actively advocating XP in the future?
31. Are you trained in UML or a similar modeling language?
32. If you know UML, did you miss it?
33. Would you like to use UML combined with an XP approach, e.g. for generation of code or tests?

**How the data was gathered**
To achieve credible results, the questionnaire was distributed among several channels worldwide. Mailing lists, direct contact and addresses of contact persons found in the internet were used. Interestingly mailing lists were relatively inefficient (only 7 of 45 answers from there). From the directly approached persons, 22% responded with a filled in questionnaire. Others responded, that they aren't allowed to officially acknowledge that they are doing XP ("Guerilla XP").

The questionnaire contains questions to be answered with free text as well as with a numeric rating. The latter are grouped and usually represented in charts. The free text questions were evaluated and (if possible) classified according to the context of the question. Some of these answers are cited below.

## 3 RESULTS OF THE SURVEY
**Some core results**
- Almost all of the projects were rated successful.
- 100% of the asked developers would reuse XP in the next project, when appropriate.
- The frequent absence of the customer was identified as high project risk.
- Problems with XP often come from "barriers in the mind": management was skeptic, company philosophy didn't allow on-site customer, developers refused pair programming.
- Most useful XP elements were common code ownership, testing and continuous integration. Most critical metaphor and on-site customer.
- As most important success factors have been mentioned: testing, pair programming and the focus of XP on the right goals.

**Potential problems with the survey**
The filled questionnaires showed a clear trend to rate the project outcome as success. Only one of 45 was rated partial success, none as failure. This may have three reasons: (1) XP is a real silver bullet, (2) developers tend to evaluate their work more positive than customers would (and we didn't have access to customers), and (3) developers from unsuccessful XP projects don't bother about XP anymore and either haven't been reached or didn't want to answer. But the high success rate clearly demonstrates that XP enables successful projects.

The second problem is that, whenever a new technology is used, the early adopters are usually higher motivated. This alone may make XP projects more successful than traditional projects, without XP itself being superior. Reasons for XP projects were among others: "personal interest" with 17,8%, "good experience in other projects" and "customer/management wanted it" with 20%. Therefore, "personal interest" was a partial motivator and thus had some influence on the survey outcome, that we unfortunately cannot quantify.

**Statistics on the participating companies**
The companies and their continents are structured as follows (the most important countries were: USA 24%,

Germany 20%, Switzerland: 13%, UK: 13%)

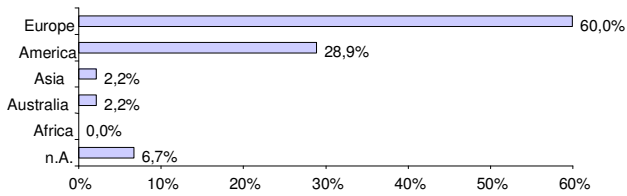

The industrial sectors split as follows:

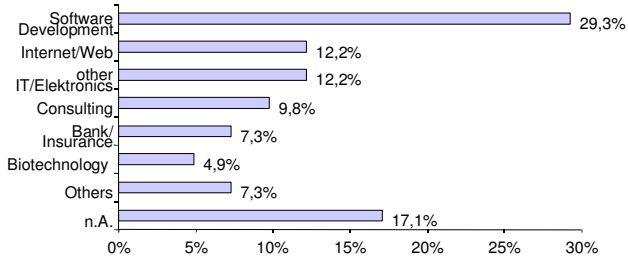

XP is used in traditional as well as new economy companies of all sizes:

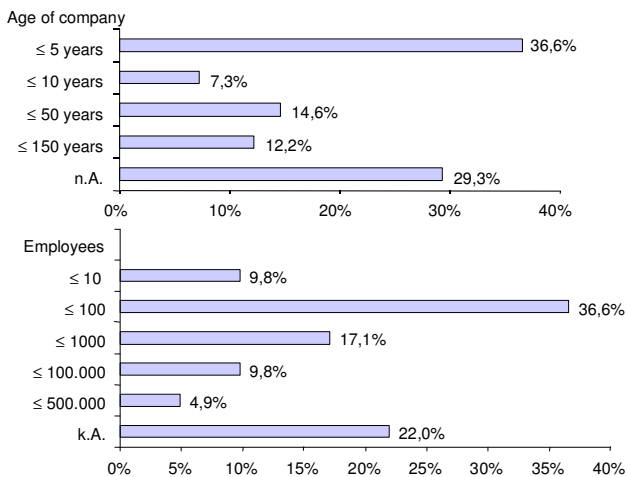

**Background of the respondents and teams**

In a third of the questioned teams the members are well experienced in software engineering in general. In another 42% of the teams the experience was mixed (experts and newcomers).

The roles of the respondents were distributed as follows:

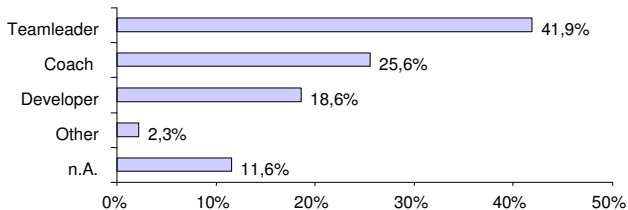

In more than 50% no external consultant was member of the project team, 21% had one consultant, 24% even more (5% didn't answer that question).

That XP has a high expansion rate can be concluded from the fact that more than half of the questionnaires were filled on the first XP project:

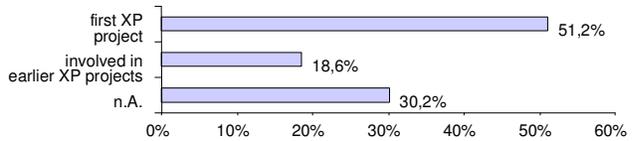

In several larger XP projects the start was quite conventional:

- *"The project itself started about two years ago using a standard development methodology. The decision to transition to XP was taken because of all the usual difficulties of managing development projects."*

**The projects**

51,1% of the projects were finished, 48,9% still running. The following project schedule indicates the rapidly growing interest in XP:

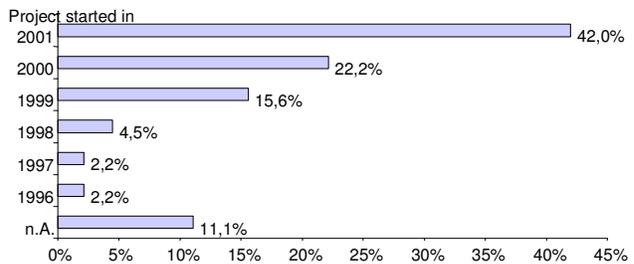

The duration of the projects was rather equally distributed among: less than six months, one year, and up to three years.

The size of the teams, however, was somewhat surprising, as larger XP projects do exist and are considered as successful:

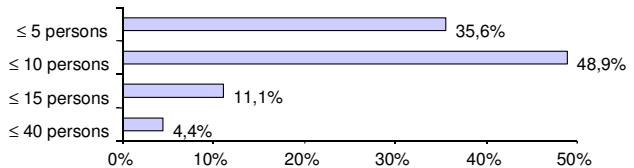

The application domain was rather mixed, with the following peaks:

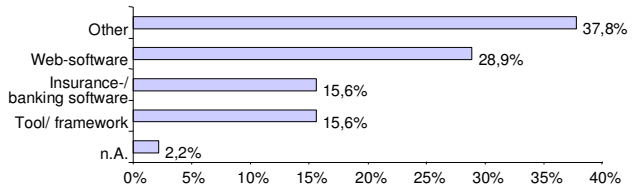

About 73% of the systems were developed completely new, the others either added new functionality to a given system (15%), developed a new part interacting with a legacy system (9%) or were maintenance projects (11%) (multiple selection was allowed).

The languages used were distributed as follows (again multiple selection allowed):

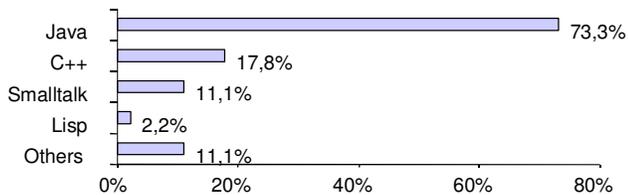

It is not surprising that XP is most efficient and therefore mainly used with high-level (OO) languages. Similar for technologies that have been used:

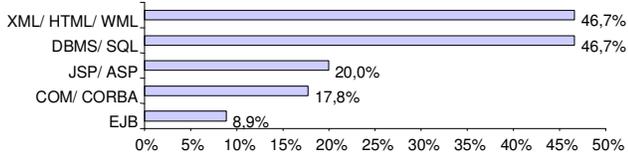

**Why XP was used?**

One of the most interesting questions: What were the reasons for applying the XP approach? The free text answers have been categorized as follows:

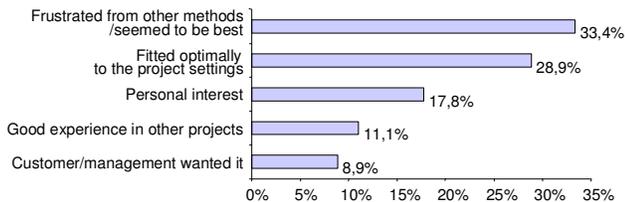

In many cases, XP seemed to be the appropriate method. Some statements:

- *"Basically, after reading and thinking and talking about it a lot, I thought that it made more sense then any other methodology I'd read about. I didn't agree with all of it but I decided we should give it a try...".*

- *"We felt that the XP is simple & better Process."*

Others were frustrated from traditional techniques and relaunched the project

- *"The project commenced in March 2000 using CMM Level 5 outsourced developers using Unified method. Code delivered unsatisfactory. Development brought in-house February 2001, and project re-started"*

**XP in the project**

55% of the projects had more than one person acting as customer, 25% at least one, 4.6% none. 14% didn't answer this question, which could lead to the assumption they had no on-site customer as well. Furthermore, several customers were "substitutes" played from the project manager, sales persons or the programmers themselves. The pretty high rate for customers indicates either that the on-site customer indeed plays a vital role in XP projects or that a higher rate of the project teams were already satisfied with a "normal" customer, who is more closely integrated into the team, but still not a perfect on-site customer.

Testers, coaches and programmers were distributed as follows:

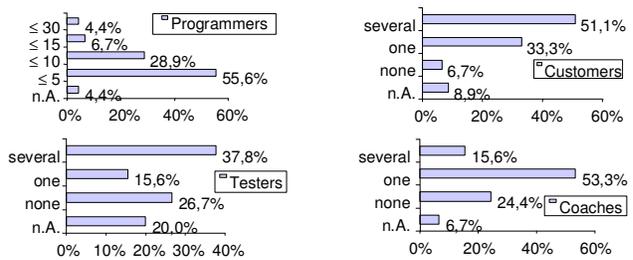

In more than 60% of the projects additional roles, such as time tracker have been mentioned.

Of particular interest have been the assessment 12 XP elements. Each of them was rated on a scale from 9 (strongly used) to 0 (not used at all). The average values and the deviation distribute as follows:

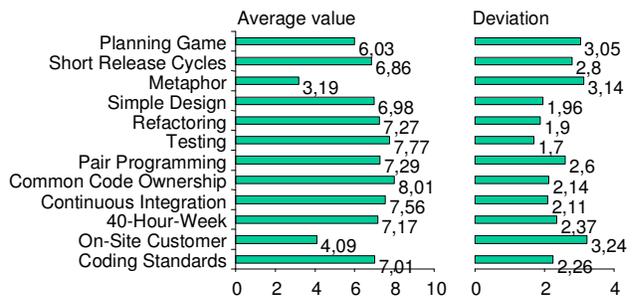

Metaphor was seen most critically: It was not used by 40% of the projects at all, because to many respondents it wasn't clear how to apply it. The on-site customer got a bad rate, mainly because customers have not been as available as it was desired. On the other hand, common code ownership seems to be the easiest to realize. So it is consistent that the metaphor and the on-site customer are the two elements that need improvement most:

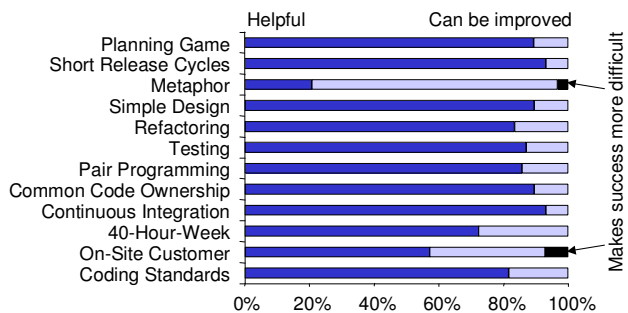

**Project goals**

Questions dealing with project progress and results were to be answered relative to traditional approaches in a scale from 5 (much better), to 0 (as always) to –5 (much worse). Interestingly none of the answers included a number below 0. This is a strong case for XP.

The detailed numbers have been split between ongoing and finished projects. The questions where, whether the costs of late changes have been reduced, the quality of the result was increased, the work was more fun, and the software can/could be delivered in time better than before:

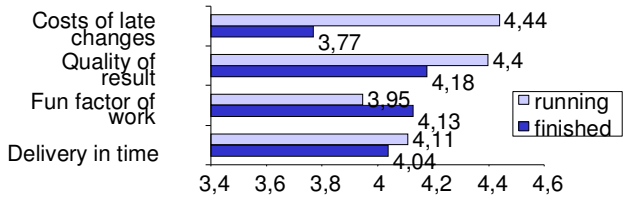

Interestingly, both the cost of change and the quality were seen less positive from the finished projects than from ongoing ones. This effect sustains that changes in later stages of the project still have higher costs of change and the projects are seen less optimistically. Furthermore, design flaws usually occur at the end of the project, thus reducing the quality ratings. But, although the optimism is less after projects are finished, the rating of 3,77 still indicates that the costs of late changes are much less in XP projects than in traditional ones. Reasons for this may be that refactoring, rigorous testing techniques and the omission of redundant documentation enables changes and lean ("simple") software produces less rework when changed. Interestingly, the fun factor for finished projects is higher than for ongoing ones. This may come the fact that people tend to forget negative experiences earlier than positive ones.

**Difficulties with XP elements**
Knowing the ratings of their usefulness, it is not surprising, what the difficult elements of the XP approach were:

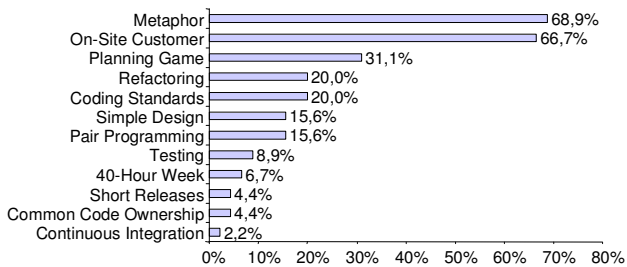

While the metaphor was not used largely, due to difficulties to understand it, the problems with the on-site customer had other reasons. Some citations:

- *"On-Site Customer: This would be great, but we did not have a chance to experience it."*
- *"Hard to convince the customer to be on-site always."*
- *"[Customer] did not participate as much as would have been preferred."*
- *"On-Site Customer, didn't use this because we couldn't get a customer to participate."*
- *"The customer was very busy on other projects ..."*

On the other hand, there were also cases, where the customer wasn't necessary all the time, or wasn't able to play his part accordingly:

- *"On-Site customer - We didn't need him on-site 100%."*
- *"Customers not really competent (or to busy) to write stories."*

This indicates how important it is to have the customer willing and able to support an XP project.

**Project success**
In a rating from 9 (full success) to 0 (failure) all except one projects rated between 7 and 9. Average of the running projects was 8,1, of the finished projects significantly smaller: 7,6. As above, this indicates that running projects are estimated more optimistic than finished ones.

The following success factors have been identified:

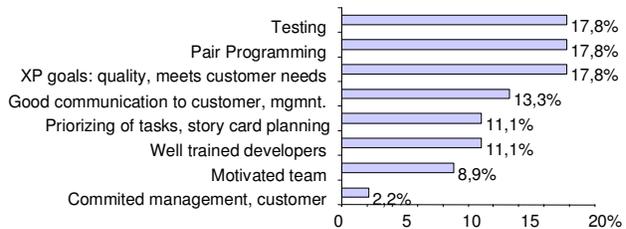

Comments on the success factors were:

- *"Tests and pair programming had prio 1 as success contributions."*
- *"Test, test, test. Write test cases first. Have a good test driver available for ALL components."*
- *"Quality Software delivered on time."*
- *"Stability and defect rate is excellent."*

Pair programming seems to be much harder to realize:

- *"I was most sceptical about this [Pair Programming] before; I'm most in favor of it now."*
- *"The Pair Programming was a major benefit to the project. Coding was completed much faster and there was immense knowledge transfer between the programmers."*
- *"Pair Programming: never decided to use it at 100\%, had two developers in team who refused to do it or were very difficult to work with."*

**Project risks**
Being asked what they consider as the most important risks for the project success, the respondents answered:

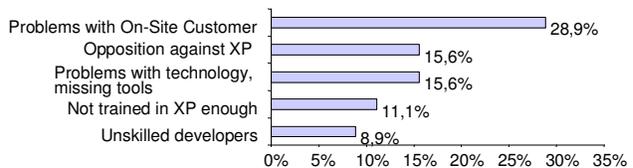

Thus the most critical problems are the missing or unwilling on-site customer, mental opposition against XP by one or some of the participants, but interestingly also technical problems. The opposition could come from a variety of sources, such as management, other departments, developers or the customer. Some comments indicate, that a non permanently available customer can be at least partially replaced by the planning game. Other comments:

- *"No customer on site. Not too dangerous since there were clear requirements and regular meetings."*
- *"Lack of an on-site customer - very dangerous, causes a lack of focus in the project."*
- *"On-site customer - the current culture of how software development "works" makes it extremely hard to apply this in practice, i.e. to involve a non-technical stakeholder as a peer within the team. Instead, the relationship between engineers and users are implicitly viewed as 'adversarial'."*
- *"On-Site Customer, difficult from a logistic point of view; not very well compatible with company's culture."*

Although, the customer(s) sometimes have been available, they caused problems by not being able to priorize tasks or to describe test plots.

Partly XP projects have been carried out without informing the customer like in a "Guerrilla XP":

- *"Project management had no trust in team and XP - very dangerous."*
- *"The only obstacle was time and the customer. The customer wasn't informed..."*

On the technological side, questions on modeling techniques such as UML showed, there is some interest in combing UML in the XP approach. 35% of the respondents used UML in the project. The desired main purpose for UML in an XP project was for communication (28,9%) and for code and test generation (13,3%). A majority of 53,3%, however, doesn't want to see UML in XP projects at all.

**Conclusions on the XP approach**

The question, whether XP shall be used again have been answered with "yes" by 93,3%, whereas the remaining 6,7% wanted an improved XP. All 100% of the respondents even want to actively advocate XP in the future. This demonstrates that XP is superior to some of the traditional approaches at least in the domains it was used. However, it also raises the question, whether, the survey only reached XP supporters and should therefore be treated carefully. This has been discussed earlier already.

### 4 OUTLOOK

This survey has to be understood as an initial survey on the use of XP in real world projects. The XP community has grown up and more and more XP projects will be finished. Therefore, more surveys on XP need to come and to refine the data gathered with our survey in Summer 2001.

Actually we feel it too early to come up with final conclusions based on this single survey, but more surveys will follow and will either strengthen or change our findings.

As discussed, the pretty high rate of developer statisfaction with XP and the equally high number of people rating their projects a success demonstrate that XP is an attractive approach to software development. The company structure also indicates, that XP is by no means restricted to the New Economy or the internet world, but is appealing for all innovative companies.

No doubt, a living methodology such as XP will improve, as the body of knowledge will grow. It will be extended with traditional elements and will be applied to new domains, such as large and well structured telecommunication systems, embedded systems software, as well as to larger projects. Furthermore, tool support will improve and lead to an adaptation of the importance of XP concepts.

Not only based on this survey, we believe XP techniques belong to the portfolio of a well trained software engineer in the same way as more traditional techniques. This enables the software engineer to flexibly react to upcoming projects needs.

**ACKNOWLEDGEMENTS**
This work was supported by the Bayerisches Staatsministerium für Wissenschaft, Forschung und Kunst through the Bavarian Habilitation Fellowship and the German Bundesministerium für Bildung und Forschung through the Virtual Softwaereengineering Competence Center (ViSEK). Special thanks go to our colleague Guido Wimmel, Michele Marchesi for helping us identifying contact persons, our partners BMW, ESG, Mummert + Partner, Siemens and all participants of the survey.